\documentclass[
 amsmath,amssymb,
superscriptaddress
groupedaddress, nofootinbib]{revtex4-2}

\usepackage{algorithm}
\usepackage{algorithmicx}
\PassOptionsToPackage{noend}{algpseudocode}
\usepackage{algpseudocode}

\makeatletter
\let\OldStatex\Statex
\renewcommand{\Statex}[1][3]{%
  \setlength\@tempdima{\algorithmicindent}%
  \OldStatex\hskip\dimexpr#1\@tempdima\relax}
\makeatother

\algnewcommand{\LineComment}[1]{\State \(\triangleright\) #1}

\usepackage{physics}

\usepackage{mathtools}

\usepackage{comment}

\usepackage{enumitem}

\usepackage{amsmath}
\newcommand{\ubf}{\boldsymbol{u}}
\newcommand{\pt}{\tilde{p}}
\newcommand{\biasV}{b}
\newcommand{\biasH}{b^\prime}

\newcommand{\weightVH}{W}
\newcommand{\dkl}{\mathcal{D}_{\mathcal{K}\mathcal{L}}}
\newcommand{\vbf}{{\boldsymbol v}}
\newcommand{\hbf}{{\boldsymbol h}}
\newcommand{\Vbf}{{\boldsymbol V}}
\newcommand{\Hbf}{{\boldsymbol H}}
\newcommand{\thetabf}{{\boldsymbol \theta}}

\newcommand{\kp}{{k^\prime}}
\newcommand{\tp}{t^\prime}
\newcommand{\xb}{{\boldsymbol x}}
\newcommand{\alb}{{\boldsymbol \alpha}}
\usepackage{amssymb}
\newcommand{\oset}{\varnothing}
\newcommand{\anglink}{{\left \langle i \right \rangle_k^n}} 
\newcommand{\angljnpk}{{\left \langle j \right \rangle_k^{n^\prime}}} 
\newcommand{\anglinkp}{{\left \langle i \right \rangle_\kp^n}} 

\begin{document}

\title{Learning Moment Closure in Reaction-Diffusion Systems \\
with Spatial Dynamic Boltzmann Distributions}

\author{Oliver K. Ernst}
\affiliation{
Department of Physics, University of California at San Diego, La Jolla, California
}

\author{Tom Bartol}
\affiliation{
Salk Institute for Biological Studies, La Jolla, California
}

\author{Terrence Sejnowski}
\affiliation{
Salk Institute for Biological Studies, La Jolla, California
}
\affiliation{
Division of Biological Sciences, University of California at San Diego, La Jolla, California
}

\author{Eric Mjolsness}
\affiliation{
Departments of Computer Science and Mathematics, and Institute for Genomics and Bioinformatics, University of California at Irvine, Irvine, California
}

\date{\today}

\begin{abstract}

Many physical systems are described by probability distributions that evolve in both time and space. 
Modeling these systems is often challenging to due large state space and analytically intractable or computationally expensive dynamics.
To address these problems, we study a machine learning approach to model reduction based on the Boltzmann machine. 
Given the form of the reduced model Boltzmann distribution, we introduce an autonomous differential equation system for the interactions appearing in the energy function.
The reduced model can treat systems in continuous space (described by continuous random variables), for which we formulate a variational learning problem using the adjoint method for the right hand sides of the differential equations.
This approach allows a physical model for the reduced system to be enforced by a suitable parameterization of the differential equations.
In this work, the parameterization we employ uses the basis functions from finite element methods, which can be used to model any physical system.
One application domain for such physics-informed learning algorithms is to modeling reaction-diffusion systems.
We study a lattice version of the R{\"o}ssler chaotic oscillator, which illustrates the accuracy of the moment closure approximation made by the method, and its dimensionality reduction power.

\end{abstract}


\maketitle


\section{Introduction}


Probability distributions that evolve in both space and time appear in many modeling applications, such as reaction-diffusion systems~\cite{hellander_2015,ramalho_2013,ruttor_2009,
bronstein_2018_pre}, neural population activities~\cite{marre_2009,odonnell_2016}, and fluid dynamics~\cite{zhao_2017}, 
as well as in engineering fields such as traffic forecasting~\cite{li_2018} and navigation of autonomous vehicles~\cite{lefevre_2014}.
However, (1) the state space of such distributions is generally large, and (2) the dynamical systems obeyed by their observables may be unknown or intractable to solve analytically. These aspects make modeling spatiotemporal systems a computational challenge, and limit the interpretability of such models.

Reaction-diffusion systems are a typical example of these problems. The distribution over system states obeys a chemical master equation (CME)~\cite{gardiner_2009}, but the state space grows exponentially with the number of random variables that describe it~\cite{munsky_2006}. Further, the time evolution of observables is not closed, i.e. the time evolution of lower order moments depends on higher order ones (similar to a BBGKY hierarchy~\cite{uhlenbeck_1963}). Their estimation therefore requires the use of a moment closure approximation (e.g.~\cite{bronstein_2018_jcp,smadbeck_2013} and others; see~\cite{johnson_2015} for a review), or otherwise sampling algorithms such as the Gillespie stochastic simulation algorithms (SSA)~\cite{gillespie_1977}, or related methods for spatial systems~\cite{mcell_1,mcell_2}.

A reduced model is one which approximates both the true distribution and its dynamics, and should address the challenges above by: (1) having a smaller state space, and (2) being more easily tractable or computationally efficient~\cite{johnson_2015}. Reduced models of reaction-diffusion systems are widely studied~\cite{hellander_2015,thomas_2012}, particularly in the context of multiscale modeling in biology~\cite{mjolsness_2018}. Recent work~\cite{ramalho_2013,bronstein_2018_pre,bronstein_2018_jcp} has demonstrated methods based on entropic matching as a highly general approach to model reduction of reaction networks.

In this paper, we demonstrate a machine learning (ML) approach to model reduction using Boltzmann machines (BM)~\cite{ackley_1985}. We formalize the methods of earlier work~\cite{johnson_2015,ernst_2018}, and extend these with the introduction of latent variables. This approach also extends work on entropic matching methods to treat spatial systems.
We present examples for spatial chemical reaction systems that demonstrate the moment closure properties of the reduced model, and apply the method to learn a spatial chaotic oscillator.

ML approaches have emerged as a powerful tool for studying quantum many-body problems~\cite{carleo_2017,han_2018}.
The area of ML most suited for model reduction of reaction-diffusion systems are generative models~\cite{mehta_2018}, where it is assumed that data are samples of an unknown probability distribution, with the goal of estimating this distribution by a structured approach.
This structure can offer insight into the problem that has not been obtainable analytically~\cite{bengio_2013}, and allows new samples to be drawn using e.g. Markov Chain Monte Carlo (MCMC) methods~\cite{mackay_2003}. Typically, a graphical model for the distribution is introduced and learned by determining interaction parameters between random variables. 

Our approach introduces an differential equation (DE) model for interaction parameters in the graph. The learning problem is formulated to determine these DEs by a maximum likelihood approach. In contrast to machine learning methods for learning temporal data such as recurrent networks, here prior information about the system may be used to enforce a reduced physical model by parameterizing the functional forms of the DEs.

A further advantage of this strategy is that it offers a natural description of systems where neither time nor space are discretized, i.e. the system is described by random variables representing space continuously and varying continuously in time. In this case, a partial differential equation (PDE) model can be introduced.
Spatially continuous descriptions are beneficial when confined geometries would introduce error into lattice-based methods, e.g. when modeling reaction-diffusion systems at synapses~\cite{mcell_1}.

The algorithmic solution to this learning problem takes the form of a PDE-constrained optimization problem. The algorithm and its derivation are closely related to BM learning, but in this case data samples are trajectories in space and time, rather than instantaneous snapshots or slices. A related framework, graph-constrained correlation dynamics~(GCCD)~\cite{johnson_2015}, has a similar learning goal, but uses spatially aggregated snapshots in time, and does not consider spatial reduced models.

The outline of this paper is: (1) in Section~\ref{sec:2} we introduce spatial dynamic Boltzmann distributions as reduced models of reaction-diffusion systems in continuous space, and formulate their learning problem using adjoint methods; (2) in Section~\ref{sec:3} we demonstrate the connection to a restricted Boltzmann machine; (3) in Section~\ref{sec:4} we demonstrate how hidden layers implement moment closure approximations, and we use the method to learn a spatial chaotic oscillator system.


\section{Spatial dynamic Boltzmann distributions} \label{sec:2}


In this section, we introduce the reduced model for a spatiotemporal distribution and its dynamics in continuous space from~\cite{ernst_2018}, and formulate the learning problem using adjoint methods. We consider the specific application of a reaction-diffusion system, but note that the methods are also applicable to other spatiotemporal systems.

The state of a reaction-diffusion system at some time $t$ is described by $n$ particles of species labels $\alb$ located at positions $\xb$ in generally continuous 3D space (each $x_i$ for $i=1,\dots,n$ is a coordinate in 3D space). Let the true distribution over system states be denoted by $p(n,\alb,\xb,t)$, which evolves in time according to the chemical master equation (CME). 

To define the reduced model, introduce $k$-particle interaction functions $\nu_k(\alb_\anglink, \xb_\anglink, t)$, where $\anglink$ denotes any ordered subset of $k$ indexes with each index in $\{ 1,\dots,n \}$. Given a set of such interaction functions $\{ \nu \}_{k=1}^K$ up to cutoff order $K$, define a \textit{spatial dynamic Boltzmann distribution} as one of the form:
\begin{equation}
\begin{split}
& \pt(n,\alb,\xb, t; \{ \nu \} ) = \\ 
& \frac{1}{Z [\{ \nu \}] } \exp [ - \sum_{k=1}^K \sum_\anglink \nu_k (\alb_\anglink, \xb_\anglink, t) ]
,
\end{split}
\label{eq:sdboltz}
\end{equation}
where the sum over $\anglink$ iterates over unique $k$-th order interactions between $n$ particles, and the partition function is
\begin{equation}
\begin{split}
& Z [\{ \nu \}] = \\ 
& \sum_{n=0}^\infty \sum_\alb \int d\xb \; \exp [ - \sum_{k=1}^K \sum_\anglink \nu_k (\alb_\anglink, \xb_\anglink, t) ]
\end{split}
\end{equation}

Boltzmann distributions are maximum entropy (MaxEnt) distributions, where each interaction function $\nu_k(\alb_\anglink,\xb_\anglink,t)$ controls a corresponding moment $\mu_k(\alb_\anglink, \xb_\anglink,t)$, given by:
\begin{widetext}
\begin{equation}
\mu_k(\alb_\anglink, \xb_\anglink,t)
= \sum_{n^\prime=0}^\infty \sum_{\alb^\prime} \int d\xb^\prime \; p(n^\prime,\alb^\prime,\xb^\prime,t)
\sum_\angljnpk \delta(\xb_\anglink - \xb_\angljnpk^\prime) \delta(\alb_\anglink - \alb_\angljnpk^\prime) 
,
\label{eq:maxentC}
\end{equation}
\end{widetext}
that is, the average number of $k$-sized tuplets of particles of species $\alb_\anglink$ at locations $\xb_\anglink$. Note that $\alb^\prime$ and $\xb^\prime$ are of size $n^\prime$. 


\subsection{Moment matching} \label{sec:2.1}


Given a set of training data drawn from $p(n,\alb,\xb,t)$ at some instant in time, the BM learning algorithm determines parameters in the energy function such that the instantaneous distribution~(\ref{eq:sdboltz}) is the MaxEnt dist. consistent with the moments in the dataset. To learn a reduced model of a system that evolves in both time and space continuously, we seek the distribution that is \textit{at all times} the MaxEnt solution. Define as the action the KL-divergence between the true and reduced models, $p$ and $\pt$, over all times:
\begin{equation}
S =\int_{t_0}^{t_f} dt \; \dkl(p || \pt)
\label{eq:action}
\end{equation}
where the Lagrangian is ${\cal L}(t ; \{ \nu \}) = \dkl(p || \pt)$ for
\begin{equation}
\begin{split}
\dkl (p || \pt) = & \sum_{n=0}^\infty \sum_\alb \int d\xb \; \\
& p(n,\alb,\xb,t) \ln \frac{p(n,\alb,\xb,t)}{\pt (n,\alb,\xb,t; \{ \nu \})} .
\end{split}
\end{equation}
Minimizing $S$ is thus equivalent to maximizing the log-likelihood of the observed data given the interaction functions, i.e. $L(\{ \nu \} ; \alb, \xb, t) = \log \pt (\alb,\xb,t ; \{ \nu \} )$. Other approaches for modeling time series are discussed in Section~\ref{sec:3.1}.

The condition for extremizing the action follows from the chain rule as
\begin{align}
\begin{split}
& \delta S
= 
\int_{t_0}^{t_f} dt \sum_{n=0}^\infty \sum_{\alb} \int d\xb \; \\
& \sum_{k=1}^K
\sum_\anglink
\Delta \mu_k(\alb_\anglink, \xb_\anglink,t)
\delta \nu_k (\alb_\anglink,\xb_\anglink,t)
= 0
,
\end{split}
\label{eq:varProblemGeneral}
\end{align}
where
\begin{equation}
\begin{split}
\Delta \mu_k(\alb_\anglink, \xb_\anglink,t)
=
& \tilde{\mu}_k(\alb_\anglink, \xb_\anglink,t) \\
& - \mu_k(\alb_\anglink, \xb_\anglink,t)
\end{split}
\label{eq:momentDiff}
\end{equation}
where $\mu$ and $\tilde{\mu}$ are averages taken over $p$ and $\tilde{p}$. This appearance of a difference of moments is the common result from using the KL-divergence in the objective functional.


\subsection{An adjoint method learning problem for spatial dynamic Boltzmann distributions} \label{sec:2.2}


Introduce for each interaction function $\nu_k (\alb_\anglink, \xb_\anglink, t)$ a \textit{functional} model:
\begin{equation}
\frac{d}{dt} \nu_k (\alb_\anglink, \xb_\anglink, t) = {\cal F}_k [ \{ \nu \} ] (\alb,\xb,t)
,
\label{eq:pdeGeneral}
\end{equation}
with initial condition $\nu_k(\alb_\anglink, \xb_\anglink, t_0) = \eta_k(\alb_\anglink, \xb_\anglink)$, and where $\{ \nu \} = \{ \nu_k \}_{k=1}^K$. 
We use ${\cal F}$ to denote a functional, allowing for example a PDE model to be introduced. Note that the arguments to the left hand side may also appear on the right, for example through a spatial derivative term $\nabla \nu_k(\alb_\anglink,\xb_\anglink,t)$.

Introduce vector notation\footnote{
In this notation, the dot product is: ${\boldsymbol a}^\intercal (\alb,\xb) {\boldsymbol b} (\alb,\xb) = \sum_{k=1}^K \sum_\anglink a(\alb_\anglink, \xb_\anglink) b(\alb_\anglink,\xb_\anglink)$.
} ${\boldsymbol \nu}(\alb,\xb,t)$ and $\boldsymbol{{\cal F}} [ \{ \nu \} ] (\alb,\xb,t)$ for the left and right hand sides of~(\ref{eq:pdeGeneral}), which contain $N = \sum_{k=1}^K \binom{n}{k}$ entries, one for every possible $(k, \anglink)$ in some order $i=1,\dots,N$. To enforce the constraint~(\ref{eq:pdeGeneral}), define the Lagrangian as the functional:
\begin{widetext}
\begin{equation}
{\cal L} [\{ \nu \}, \{ \xi \}] (t) = \dkl (p || \pt) 
+ 
\sum_{n=0}^\infty \sum_\alb \int d \xb \; \boldsymbol{\zeta}^\intercal (\alb, \xb, t) \left ( \frac{d \boldsymbol{\nu}(\alb, \xb, t)}{dt} - \boldsymbol{{\cal F}} [ \{ \nu \} ] (\alb,\xb,t) \right )
\label{eq:lagrangian}
\end{equation}
\end{widetext}
where we have introduced Lagrange multiplier functions $\boldsymbol{\zeta}(\alb, \xb, t)$ corresponding to $\boldsymbol{\nu}(\alb, \xb, t)$. Since the constraint is satisfied, then the action is as before~${S = \int_{t_0}^{t_f} dt \; {\cal L} [\{ \nu \}, \{ \xi \}] (t)}$ where $\{ \xi \} = \{ \xi_k \}_{k=1}^K$.  

Introducing perturbations $\delta \boldsymbol{\nu}(\alb, \xb, t)$ to the interaction functions gives as condition for extremizing the action:
\begin{widetext}
\begin{equation}
\delta S = \int_{t_0}^{t_f} dt \; 
\sum_{n=0}^\infty \sum_\alb \int d \xb \;
\delta \boldsymbol{\nu}^\intercal (\alb,\xb,t)
\Bigg \{
\Delta \boldsymbol{\mu} (\alb,\xb,t) 
- \frac{d \boldsymbol{\zeta}(\alb, \xb, t)}{dt}
- \frac{\delta {\cal J}[\{ \nu \}, \{ \zeta \}] (t)}{\delta \boldsymbol{\nu}(\alb, \xb, t)} 
\Bigg \} 
= 0
,
\label{eq:varyS}
\end{equation}
\end{widetext}
where the boundary terms from the integration by parts in the second term have vanished due to the boundary condition for the adjoint variables $\boldsymbol{\zeta} (\alb, \xb, t_f) = 0$, and we have defined:
\begin{equation}
\begin{split}
{\cal J}[\{ \nu \}, \{ \zeta \}] (t) 
=
\sum_{n^\prime=0}^\infty & \sum_{\alb^\prime} \int d \xb^\prime \; \\
& \boldsymbol{\zeta}^\intercal (\alb^\prime, \xb^\prime, t)
\boldsymbol{{\cal F}} [ \{ \nu \} ] (\alb^\prime, \xb^\prime,t)
\end{split}
\end{equation}
We therefore obtain the adjoint system
\begin{equation}
\frac{d \boldsymbol{\zeta}(\alb, \xb, t)}{dt}
=
\Delta \boldsymbol{\mu} (\alb,\xb,t)
- \frac{\delta J[\{ \nu \}, \{ \zeta \}] (t) }{\delta \boldsymbol{\nu}(\alb, \xb, t)} 
.
\label{eq:adjointDE}
\end{equation}
Depending on the form of the functional, additional boundary conditions may be enforced to evaluate the term on the right. Equations~(\ref{eq:pdeGeneral},\ref{eq:adjointDE}) can be equivalently expressed by the Hamiltonian system
\begin{equation}
\begin{split}
\frac{d \boldsymbol{\nu}(\alb,\xb,t)}{dt} &= \frac{\delta H [ \{ \nu \}, \{ \zeta \} ] (t) }{\delta \boldsymbol{\zeta}(\alb,\xb,t)} , \\
\frac{d \boldsymbol{\zeta}(\alb,\xb,t)}{dt} &= - \frac{\delta H [ \{ \nu \}, \{ \zeta \} ] (t) }{\delta \boldsymbol{\nu}(\alb,\xb,t)} ,
\end{split}
\end{equation}
where
\begin{equation}
H [ \{ \nu \}, \{ \zeta \} ] (t) = - \dkl (p || \pt) + {\cal J}[\{ \nu \}, \{ \zeta \}] (t) .
\end{equation}

Given a reduced model for the dynamics~(\ref{eq:pdeGeneral}), equation~(\ref{eq:varyS}) gives the necessary condition for extremizing the action. In a typical model reduction setting, however, the reduced model is not know beforehand. What should the form of the model~(\ref{eq:pdeGeneral}) be extremize the action? Consider the case where the functional is specified in terms of some ordinary functions. We next set up a variational problem for these functions appearing on the right hand side of the differential equation. Variational problems of this form have been studied previously: first in the context of optimal control theory ~\cite{gamkrelidze_1967,neustadt_1970}, and later didactically in~\cite{gamkrelidze_book}. 

Let the functional be of the form:
\begin{equation}
\frac{d}{dt} \nu_k (\alb_\anglink, \xb_\anglink, t) = {\cal F}_k [ \{ \nu \}, \{ F_k \} ] (\alb,\xb,t)
,
\label{eq:pdeGeneralOrdinary}
\end{equation}
where the $M_k$ ordinary functions appearing on the right hand side are $F_k^{(s)}(\{ \nu(\alb,\xb,t) \})$ for $s=1,\dots,M_k$, denoted by $\{ F_k \} = \{ F_k^{(s)} \}_{s=1}^{M_k}$. For arbitrary perturbations $\delta F_k^{(s)}$, extremizing the action gives
\begin{widetext}
\begin{equation}
\delta S = - \int_{t_0}^{t_f} dt \; 
\sum_{n=0}^\infty \sum_\alb \int d \xb \; \sum_{k=1}^K \sum_\anglink \sum_{s=1}^{M_k}
\frac{\delta {\cal J}[\{ \nu \}, \{ \zeta \}] (t) }{\delta F_k^{(s)} (\{ \nu(\alb,\xb,t) \})} 
\delta F_k^{(s)} (\{ \nu(\alb,\xb,t) \})
= 0
\label{eq:varProblem}
\end{equation}
\end{widetext}

Equation~(\ref{eq:varProblem}) is the variational calculus form of the sensitivity equation obtained by the adjoint method when the functional model is specified in terms of some parameter vector~\cite{cao_2003}. This is particularly clear if we consider the specific form of~(\ref{eq:pdeGeneralOrdinary}) as the autonomous ordinary differential equation (ODE) system:
\begin{equation}
\frac{d}{dt} \nu_k (\alb_\anglink, \xb_\anglink, t) = F_k (\{ \nu (\alb_\anglink, \xb_\anglink, t) \} )
,
\label{eq:pdeAutonomous}
\end{equation}
where $\{ \nu (\alb_\anglink, \xb_\anglink, t) \}$ denotes all $\nu$ of all possible arguments appearing on the left hand side. In this case,~(\ref{eq:varProblem}) becomes
\begin{equation}
\begin{split}
& \delta S = \\
& - \int_{t_0}^{t_f} dt \; 
\sum_{n=0}^\infty \sum_\alb \int d \xb \;
\boldsymbol{\zeta}^\intercal (\alb, \xb, t) 
\delta \boldsymbol{F} (\{ \nu (\alb, \xb, t) \})
= 0
\label{eq:varProblemAutonomous}
,
\end{split}
\end{equation}
where as before we have used vectors of length $N$ to denote possible $(k,\anglink)$ as before. This resembles the adjoint method sensitivity equation, where variational terms $\delta F_k$ and $\delta S$ replace ordinary derivatives with respect to parameters. This will be pursued further in Section~\ref{sec:3.1}. The result of~(\ref{eq:varProblemAutonomous}) is that extremizing the action requires that the adjoint variables vanish everywhere $\zeta_k (\alb_\anglink,\xb_\anglink,t) = 0$. One case when this is satisfied is if the adjoint system is source free $\Delta \mu_k (\alb_\anglink, \xb_\anglink, t) = 0$, i.e. the moment matching condition is enforced. 

From the Euler-Lagrange equations~(\ref{eq:adjointDE}), the adjoint variables obey:
\begin{equation}
\frac{d {\boldsymbol \zeta } (\alb,\xb,t)}{dt} = \Delta {\boldsymbol \mu}(\alb,\xb,t) - G^\intercal(\alb,\xb,t) {\boldsymbol \zeta } (\alb,\xb,t)
,
\label{eq:adjointDEautonomous}
\end{equation}
where the elements of the $N\times N$ matrix $G$ are
\begin{equation}
G_{i,i^\prime}(\alb,\xb,t) = \frac{
\partial F_k(\{ \nu(\alb_\anglink,\xb_\anglink,t) \})
}{
\partial \nu_\kp (\alb_\anglinkp,\xb_\anglinkp,t)
}
,
\end{equation}
where $(k,\anglink)$ corresponds to index $i$ and $(\kp,\anglinkp)$ corresponds to index $i^\prime$. Appendix~\ref{app:1} gives the formal solution to~(\ref{eq:adjointDEautonomous}) and makes explicit the connection between the conditions for extrema~(\ref{eq:varProblemAutonomous}) and~(\ref{eq:varProblemGeneral}).


\section{Dynamics for Restricted Boltzmann machines} \label{sec:3}


We next consider a specific case of the formalism of Section~\ref{sec:2} where the system is described by discrete random variables. A Boltzmann distribution on a state $\vbf = \{ v_1, \dots, v_N \}$ of $N$ discrete random variables
is of the form:
\begin{equation}
\pt(\vbf) = \frac{1}{Z} \exp [ - E (\vbf) ] 
,
\end{equation}
where $Z$ is the partition function, and the energy function $E(\vbf)$ is typically defined by a chosen Markov random field (MRF). For example, a Boltzmann machine (BM)~\cite{ackley_1985} is a binary MRF, where binary units update their state based on a bias and pairwise connections to other units. A MRF where all variables $\vbf$ are driven by data is fully visible; 
otherwise units $\hbf$ which are not driven by data are denoted as hidden. 

A restricted Boltzmann machine (RBM)~\cite{smolensky_1986} is a BM in which hidden and visible units are organized into layers, where a layer is defined by the property that there are no interactions among units in the same layer. For example, a typical energy function for an RBM is of the form:
\begin{equation}
E (\vbf, \hbf, \thetabf ) = - \sum_{i} \biasV_i v_i - \sum_{j} \biasH_j h_j 
- \sum_{\{ i, j \}} \weightVH_{i,j} v_{i} h_{j}
,
\label{eq:bmGeneral}
\end{equation}
where the summation $\{ i,j \}$ is determined by the graph edges, and $\thetabf$ is the vector of length $K$ of all interaction parameters in the graph. This defines a joint distribution over $\vbf$ and $\hbf$:
\begin{equation}
\pt(\vbf,\hbf ; \thetabf ) = \frac{1}{Z(\thetabf)} \exp [ - E (\vbf, \hbf, \thetabf ) ]
.
\label{eq:BMstationary}
\end{equation}
Each parameter $\theta_k$ in this MaxEnt distribution controls a corresponding moment $\tilde{\mu}_k$, given by
$\tilde{\mu}_k = \partial \ln Z(\thetabf) / \partial \theta_k$.

Define a \textit{dynamic Boltzmann distribution} as one with time-dependent interaction parameters:
\begin{equation}
\pt(\vbf,\hbf ; \thetabf(t)) = \frac{1}{Z(\thetabf(t))} \exp [ - E (\vbf, \hbf , \thetabf(t) ) ]
.
\end{equation}
For example, the energy function of the RBM becomes:
\begin{equation}
\begin{split}
E (\vbf, \hbf, \thetabf(t) ) = & - \sum_{i} \biasV_i(t) v_i - \sum_{j} \biasH_j(t) h_j  \\
& - \sum_{\{ i, j \}} \weightVH_{i,j}(t) v_{i} h_{j}
.
\end{split}
\label{eq:bmDynamic}
\end{equation} 
This is a specific case of a spatial dynamic Boltzmann distribution~(\ref{eq:sdboltz}) in the discrete lattice limit. To see this, assign to every visible unit $v_i$ a spatial location $x_i$. By taking self interaction functions $\nu_1(x,t) = \sum_i b_i(t) \delta_{x,x_i}$ in~(\ref{eq:sdboltz}), we recover the first term in~(\ref{eq:bmDynamic}) with $v_i \in \{ 0,1 \}$, where $\delta_{x,x_i}$ is unity if the coordinates are coincident and zero otherwise.

Similarly, hidden units can also be represented in continuous space. Let the species labels $\alpha_v$ denote visible units and $\beta_h$ denote hidden units, and assign to every hidden unit $h_j$ a spatial location $y_j$. The weights between layers are then obtained by taking pairwise interactions $\nu_2(x,y,\alpha,\beta,t) = \sum_{i,j} W_{i,j}(t) \delta_{x,x_i} \delta_{y,y_j} \delta_{\alpha,\alpha_v} \delta_{\beta,\beta_h}$ in~(\ref{eq:sdboltz}).


\subsection{An adjoint method learning problem for restricted Boltzmann machines} \label{sec:3.1}


\begin{figure*}[t]
\begin{minipage}{\linewidth}
\begin{algorithm}[H] 
\caption{Stochastic Gradient Descent for Learning 
Restricted Boltzmann Machine Dynamics} \label{alg:1}
\begin{algorithmic}[1]
\State \textbf{Initialize}
\State \hspace{1.4em}Parameters $\ubf_k$ controlling the functions $F_k(\thetabf ; \ubf_k)$ for all $k=1,\dots,K$.
\State \hspace{1.4em}Time interval $[t_0,t_f]$, a formula for the learning rate $\lambda$.
\While{not converged}
	\State Initialize $\Delta F_{k,i} = 0$ for all $k=1,\dots,K$ and parameters $i = 1,\dots,M_k$. \;
    \For{sample in batch}
		\LineComment{\textit{Generate trajectory in reduced space $\thetabf$:}}
		\State Solve the PDE constraint~(\ref{eq:pdeParameterized}) for $\theta_k(t)$ with a given IC $\theta_{k,0}$ over $t_0 \leq t \leq t_f$, for all $k$. \;
		\LineComment{\textit{Awake phase:}}
        \State Evaluate moments $\mu_k(t)$ of the data for all $k,t$. \;
        \LineComment{\textit{Asleep phase:}}
        \State Evaluate moments $\tilde{\mu}_k(t)$ of the Boltzmann distribution. \;
        \LineComment{\textit{Solve the adjoint system:}}
		\State Solve the adjoint system~(\ref{eq:adjoint}) for $\phi_k(t)$ for all $k,t$.
        \LineComment{\textit{Evaluate the objective function:}}
        \State Update $\Delta F_{k,i}$ as the cumulative moving average of the sensitivity equation~(\ref{eq:sensitivity}) over the batch. \;
	\EndFor
    \LineComment{\textit{Update to decrease objective function:}}
	\State $u_{k,i} \rightarrow u_{k,i} - \lambda \Delta F_{k,i}$ for all $k,i$. \;
\EndWhile
\end{algorithmic}
\end{algorithm}
\end{minipage}
\end{figure*}

Introduce for each interaction parameter $\theta_k$, $k=1,\dots,K$, in the interaction graph a \textit{time-evolution function} $F_k$ forming an autonomous ODE system (analogous to~(\ref{eq:pdeAutonomous})):
\begin{equation}
\frac{d}{dt} \theta_k(t) = F_k (\thetabf(t) )
,
\label{eq:pde}
\end{equation}
with initial condition $\theta_k(t_0) = \theta_{k,0}$. To obtain from the variational problem derived in Section~\ref{sec:2.2} an ordinary optimization problem for parameters, further consider the paramaterization by the vectors $\ubf_k$ of size $M_k$, generally unique for every $k$:
\begin{equation}
\frac{d}{dt} \theta_k(t) = F_k (\thetabf(t) ; \ubf_k )
.
\label{eq:pdeParameterized}
\end{equation}

Analogously to the continuous case, define as the objective function the KL-divergence between the true and reduced models, $p$ and $\pt$, over all times (analogous to~(\ref{eq:action})):
\begin{equation}
\begin{split}
S =& \int_{t_0}^{t_f} dt \; \dkl(p || \pt) \\ 
\dkl (p || \pt) =& \sum_{\boldsymbol z} p({\boldsymbol z}) \ln \frac{p({\boldsymbol z})}{\pt ({\boldsymbol z} ; \{ \ubf \} )} 
.
\end{split}
\label{eq:actionDiscrete}
\end{equation}
where $\{ \ubf \} = \{ \ubf_k \}_{k=1}^K$. Minimizing $S$ is thus equivalent to maximizing the log-likelihood of the observed data given the parameters, i.e. $L(\{ \ubf \} ; {\boldsymbol z}) = \log \pt ({\boldsymbol z} ; \{ \ubf \} )$. A more common approach is to instead maximize the conditional likelihood of observations conditioned on the first observation: $L(\{ \ubf \} ; z_2, z_3, \dots | z_1 ) = \log \pt ( z_2, z_3, \dots | z_1 ; \{ \ubf \} )$, or similar causal relations. For Markov chains, this approach is highly successful (leading to e.g. Kalman filters; see~\cite{ghahramani_1998} for an introduction). If a prior is available, Bayesian methods that compute the posterior $\pt (\{ \ubf \} ; {\boldsymbol z}) \propto \pt ({\boldsymbol z} ; \{ \ubf \} ) \times \pt(\{ \ubf \})$ can provide further improvements. The advantage of the current approach is that a reduced physical model can be enforced through the parameterization~(\ref{eq:pdeParameterized}). This model can be based on prior information, such as reaction networks with known solutions~\cite{ernst_2018}. A second advantage is that the generalization to spatially continuous systems follows naturally using PDEs as in~(\ref{eq:pdeGeneral}).

The time integral in $S$ can be lead to undesired extrema, for example for periodic systems where the objective function may not minimize the KL-divergence pointwise. One algorithmic strategy for eliminating these in practice is to shift the limits of integration during the optimization, as further explored in Section~\ref{sec:4.1}.

Minimizing the objective function defines a PDE-constrained optimization problem: minimize~(\ref{eq:actionDiscrete}) subject to the PDE-constraint~(\ref{eq:pdeParameterized}). Define the Lagrangian function (analogous to~(\ref{eq:lagrangian})):
\begin{equation}
\begin{split}
{\cal L}(t ; \{ \ubf \} ) =& \dkl(p||\pt) \\
&+ \sum_{k=1}^K \phi_k(t) \left ( \frac{d}{dt} \theta_k(t) - F_k (\thetabf(t) ; \ubf_k ) \right )
\end{split}
\end{equation}
where we have introduced the adjoint variables $\phi_k$ associated with each $\theta_k$. Taking the derivative of the objective function $S = \int_{t_0}^{t_f} dt \; {\cal L} (t ; \{ \ubf \})$ with respect to a parameter gives the sensitivity equation (analogous to~(\ref{eq:varProblemAutonomous})):
\begin{equation}
\frac{d S}{d u_{k,i}} = - \int_{t_0}^{t_f} dt \; \frac{\partial F_k (\thetabf(t) ; \ubf_k)}{\partial u_{k,i}} \phi_k(t)
\label{eq:sensitivity}
\end{equation}
and taking the derivative with respect to $\theta$ gives the ODE system obeyed by the adjoint variables (analogous to~(\ref{eq:adjointDEautonomous})):
\begin{equation}
\frac{d}{dt} \phi_k(t) =
\tilde{\mu}_k (t) - \mu_k (t)  
- \sum_{l=1}^K \frac{\partial F_l (\thetabf(t) ; \ubf_l)}{\partial \theta_k(t)} \phi_l(t)
\label{eq:adjoint}
\end{equation}
where $\mu_k(\tp)$ and $\tilde{\mu}_k(\tp)$ are averages taken over to $p$ and $\pt$ at time $\tp$, and the boundary condition is $\phi_k(t_f) = 0$. 

Algorithm~\ref{alg:1} outlines how this optimization problem can be solved in practice. The inner loop of an ``awake" and ``asleep" phase of sampling are identical to that of BM learning. Standard algorithmic improvements are possible, such as the use of accelerated gradient descent methods such as Adam~\cite{adam}, and using persistent contrastive divergence (PCD) method~\cite{tieleman_2008} to estimate the moments of the reduced model $\tilde{\mu}_k(\tp)$.

Adjoint methods such as these for solving PDE-constrained optimization problems are also called ``black-box" methods~\cite{herzog_2010,funke_2013}, since the PDE constraint~(\ref{eq:pdeParameterized}) is eliminated in the derivation of the sensitivity equation~(\ref{eq:sensitivity}). A competing class of methods (sometimes referred to as ``all-at-once" methods) treat the constraint explicitly in the optimization, and may offer a computational advantage over this approach. These include sequential quadratic programming (SQP) and augmented Lagrangian methods. 

Additional constraints or regularization terms can be included in the optimization, such as conserved quantities identified from the left null space of the net stoichiometry matrix. 
For example, $L_2$ regularization can be incorporated into the objective function:
\begin{equation}
S = \int_{t_0}^{t_f} dt \; \dkl(p || \pt) + \lambda_r \int_{t_0}^{t_f} dt \; \sum_{k=1}^K \left ( \theta_k(t) - \overline{\theta}_k(t) \right )^2
,
\label{eq:l2Reg}
\end{equation}
where $\overline{\theta}_k(t)$ are some specified functions or otherwise constant, and $\lambda_r$ is a regularization parameter. In this case, the adjoint variables are given by:
\begin{equation}
\begin{split}
\frac{d}{dt} \phi_k(t) =
& \tilde{\mu}_k (t) - \mu_k (t) + 2 \lambda_r \left ( \theta_k(t) - \overline{\theta}_k(t) \right ) \\
& - \sum_{l=1}^K \frac{\partial F_l (\thetabf(t) ; \ubf_l)}{\partial \theta_k(t)} \phi_l(t)
.
\end{split}
\end{equation}


\subsection{Finite element parameterization} \label{sec:3.2}


What choice should be made for the parameterization~(\ref{eq:pdeParameterized}) of the right hand sides of the differential equations? In~\cite{ernst_2018}, we considered simple reaction-diffusion systems from which general forms of approximate models could be inferred that maintain physical interpretations. A second approach also explored in~\cite{ernst_2018} is to use a separate moment closure approximation to derive analytic solutions for simple reaction systems on 1D lattices, where the inverse Ising problem is analytically solvable. The form of~(\ref{eq:pdeParameterized}) can then be taken as either linear or non-linear combinations of known solutions.

Here, we take a \textit{finite element method} (FEM)~\cite{hughes_2000} approach to the parameterization that is more aligned with the unsupervised learning problem in a Boltzmann machine. The space of solutions to the general variational problem~(\ref{eq:varProblem}), which is some Banach space, is therefore restricted to the space of finite element method solutions.

An important restriction is that the learning rule~(\ref{eq:sensitivity}) requires $C_1$ finite elements. One choice for such elements is the $Q_3$ family of finite elements~\cite{fem_table}, which in dimensions higher than one are easily constructed as tensor products of 1D cubic polynomials\footnote{
An alternative choice for tetrahedral meshes is the $P_3$ family of finite elements.
}.
For $C_1$ elements that control the value of the function and its derivative at the endpoints, these polynomials are just the Hermite polynomials, shown in Figure~\ref{fig:bimolModels}(d).

We therefore introduce for each time-evolution function in~(\ref{eq:pdeParameterized}) a domain of hypercubic cells, with $4^{d}$ degrees of freedom, where $d$ are the number of arguments to $F_k$. In practice, we found it is rarely necessary to have more than $d=3$ arguments (see Section~\ref{sec:4}). For $d=3$, each cube has 64 degrees of freedom (8 degrees of freedom at each vertex, specifying the function value and derivatives). For a cubic lattice of $V = L_1 \times L_2 \times L_3$ cells, there are $8 V$ degrees of freedom total, with the parameterization taking the usual form in terms of the basis functions $f_l$ associated with each degree of freedom:
\begin{equation}
F_k (\theta_1, \theta_2, \theta_3; \ubf_k) = \sum_{l=1}^{8 V} u_l f_l(\theta_1, \theta_2, \theta_3)
\label{eq:q3}
\end{equation}
Note that here, the right hand side of the differential equation is parameterized (as opposed to the solution of the differential equation), since the objective of the learning algorithm is to determine a suitable differential equation model.


\section{Learning reaction-diffusion systems on lattices} \label{sec:4}


Recall that the state of a reaction-diffusion system at some time is described by $n$ particles of species $\alb$ located at positions $\xb$ in generally continuous 3D space. To make an explicit connection to binary random variables, we consider a simpler model of particles hopping on a discrete lattice in the single-occupancy limit. To generate stochastic simulations of such a system, we adapt the method of~\citet{takayasu_1992} for a lattice-based variant of the popular Gillespie stochastic simulation algorithm (SSA)~\cite{gillespie_1977} as follows: at each timestep:
\begin{enumerate}[nosep]
\item Perform unimolecular reactions following the standard Gillespie SSA.
\item Iterate over all particles in random order; for each:
\begin{enumerate}[nosep]
\item Hop to a neighboring site, chosen at random with equal probability.
\item If the site is unoccupied, the move is accepted. If the site is occupied, a bimolecular reaction occurs with some probability; else, the move is rejected and the particle is returned to the original site.
\end{enumerate}
\end{enumerate}

The lattice on which particles hop is designated as the visible part of the MRF. Assign a unique index $i$ to each of the $N$ sites in the lattice, and let the vector of possible species be ${\boldsymbol s}$ of size $M$ in some arbitrary ordering (excluding $\oset$ to denote an empty site). Spins at a site $i$ are now multinomial units, represented as a vector $\vbf_i$ of length $M$ where entries $v_{i,\alpha} \in \{0,1\}$ for $\alpha = 1, \dots, M$ denote the absence or presence of a particle of species $s_\alpha$ (an $n$-vector model in statistical mechanics). The single-occupancy limit corresponds to the implicit constraint that the vectors are of unit length, i.e. $\sum_{\alpha=0}^M v_{i,\alpha} =1$, where $\alpha=0$ denotes an empty site. The matrix $\Vbf$ of size $N\times M$ describes the state of the visible part of the MRF, where each row denotes a lattice site.

Likewise introduce hidden layer species ${\boldsymbol s}^\prime$ of size $M^\prime$, which may be different from ${\boldsymbol s}$. Indexing all hidden sites as $j=1,\dots,N^\prime$, hidden unit vectors are $\hbf_j$ of length $M^\prime$. The state of the hidden units is $\Hbf$ of size $N^\prime \times M^\prime$, with the single occupancy constraint as before. 

The dynamic Boltzmann distribution becomes: $\pt(\Vbf,\Hbf | \thetabf(t)) = \exp [ - E (\Vbf, \Hbf, \thetabf(t) ) ] / Z(\thetabf(t))$, where interaction parameters $\thetabf(t)$ may also be species-dependent (excluding $\oset$). For example, the energy function for the RBM becomes:
\begin{equation}
\begin{split}
& E (\Vbf, \Hbf, \thetabf(t) ) =
- \sum_{i=1}^N \sum_{\alpha=1}^M \biasV_{i,\alpha}(t) v_{i,\alpha} \\
& - \sum_{j=1}^{N^\prime} \sum_{\beta=1}^{M^\prime} \biasH_{j,\beta}(t) h_{j,\beta}
- \sum_{\{ i, j \}} \sum_{\alpha,\beta} \weightVH_{i,j,\alpha,\beta}(t) v_{i,\alpha} h_{j,\beta}
.
\end{split}
\end{equation}


\subsection{Learning hidden layers for moment closure} \label{sec:4.1}


A typical problem in many-body systems is the appearance of a hierarchy of moments, where the time-evolution of a given moment depends on higher order moments. Moment closure approximations terminate this infinite hierarchy at some finite order. In this section, we develop the perspective of the learning problem~(\ref{eq:sensitivity}) as a closure approximation using a simple pedagogical example. We note some similarity to previously proposed closure schemes~\cite{smadbeck_2013,johnson_2015}, as well as to entropic matching~\cite{bronstein_2018_jcp}, although the current approach differs in the objective function~(\ref{eq:actionDiscrete}) and the formulation for spatially continuous systems in Section~\ref{sec:2}.

\begin{figure*}[!t]
	\centering
	\includegraphics[width=0.8\textwidth]{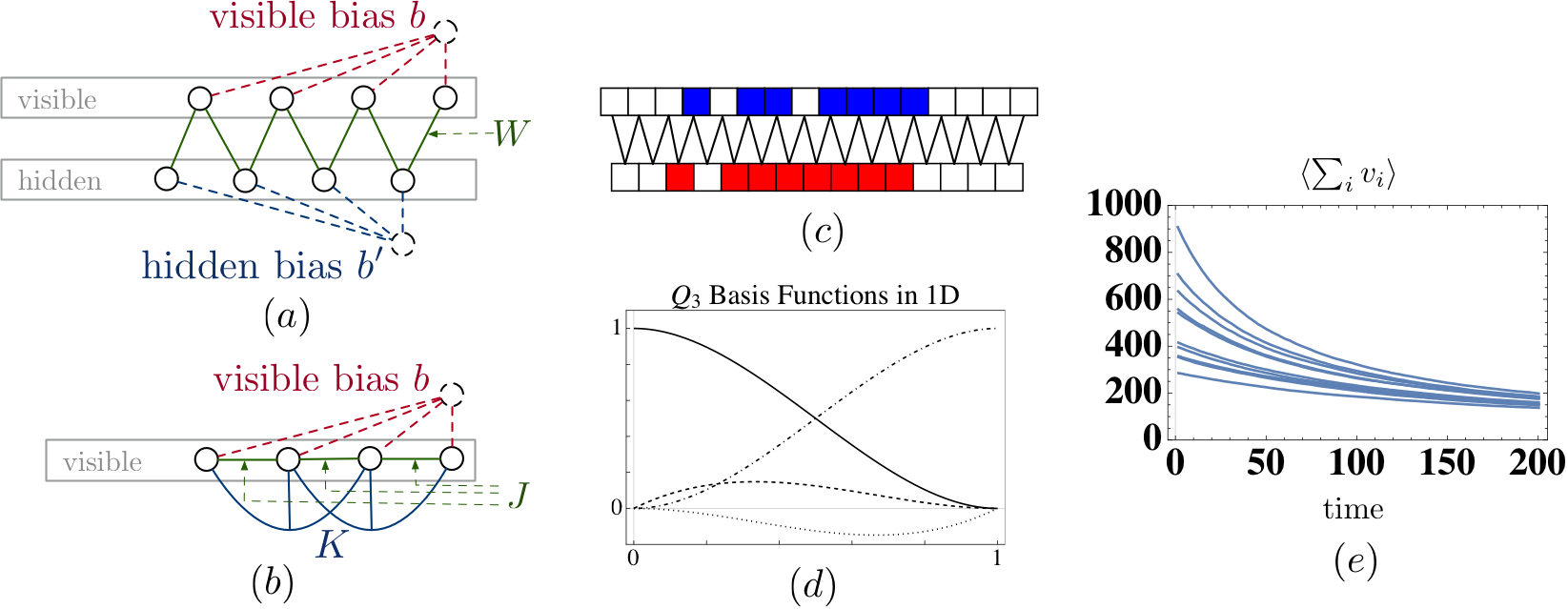}
	\caption{
	Comparison of a fully visible and a latent variable model for capturing local correlations in a 1D lattice.
	(a) 1D lattice with one hidden layer (similar to an RBM). Note that in this simplified example, $W$ is a single translation invariant parameter rather than a matrix as common in RBMs.
	(b) Fully visible model for a 1D lattice including nearest neighbor (NN) interactions $J$ and next-nearest neighbors (NNN) $K$.
	(c) An example state of the hidden layer model, where blue color indicates the presence of a particle in the visible layer, and likewise red for the hidden layer. By learning the parameters, the hidden layer can be tuned to capture the presence of NNs.
	(d) The basis functions of the $Q_3$ family of $C_1$ finite elements in 1D (Hermite polynomials), used to parameterize the right-hand sides of~(\ref{eq:model12},\ref{eq:model22}). Basis functions in higher dimensions are constructed as tensor products of the 1D polynomials. 
	(e) Moments of stochastic simulations for 10 of the 50 initial conditions used for training (each trajectory obtained from averaging over 50 lattices simulated from the same initial condition).
    }
	\label{fig:bimolModels}
\end{figure*}

Consider a bimolecular-annihilation process on a 1D lattice of length $N$, where particles of a single species $A$ hop and react according to $A + A \rightarrow \oset$. The time-evolution of the first two moments are:
\begin{equation}
\begin{split}
\frac{d}{dt} \left \langle \sum_i v_i \right \rangle =& - 2 k_r \left \langle \sum_i v_i v_{i+1} \right \rangle 
,
\\
\frac{d}{dt} \left \langle \sum_i v_i v_{i+1} \right \rangle =& 
2 D \left \langle \sum_i v_i v_{i+2} \right \rangle \\
& \hspace{-30mm} - 2 k_r \left \langle \sum_i v_i v_{i+1} v_{i+2} \right \rangle
+ ( k_r - 2D ) \left \langle \sum_i v_i v_{i+1} \right \rangle
,
\end{split}
\label{eq:momentClosure}
\end{equation}
where $k_r$ is the reaction rate and $D$ the diffusion rate. The simplest graph to capture such observables is a fully visible Markov random field, i.e. a 1D Ising model including interactions up to some order. For example, including third order interactions, let:
\begin{equation}
\begin{split}
& E ({\boldsymbol v} , b(t), J(t), K(t) ) 
= - b(t) \sum_{i=1}^N v_i \\
& - J(t) \sum_{i=1}^{N-1} v_i v_{i+1} 
- K(t) \sum_{i=1}^{N-2} v_i v_{i+1} v_{i+2} ,
\end{split}
\label{eq:model11}
\end{equation}
and let:
\begin{equation}
\begin{split}
\dot{b} =& F_b (b,J,K ; \ubf_b) , \\
\dot{J} =& F_J (b,J,K ; \ubf_J) , \\
\dot{K} =& F_K (b,J,K ; \ubf_K) ,
\end{split}
\label{eq:model12}
\end{equation}
for some parameter vectors $\ubf$ to be learned, where time derivatives are denoted as $\dot{x} = d/dt$. The corresponding graphical model is illustrated in Figure~\ref{fig:bimolModels}(b). The choice of the energy function in~(\ref{eq:model11}) defines which moments are explicitly captured by the reduced model. The additional choice of the form of the differential equations $F_\gamma$ defines the moment closure approximation made.

We next show through computational experiments that the introduction of hidden layers can improve upon a fully visible closure model:
\begin{enumerate}[nosep]
\item In any closure scheme, moments beyond a certain order are not captured explicitly by the model, so that their approximation may be poor. The representation power of hidden layers~\cite{bengio_2013} can be used to incorporate information about which higher order moments are relevant to the dataset.
\item Two distinct states having the same lower order moments are indistinguishable in the reduced model (the model is not sufficiently high dimensional). Hidden layers may be able to separate such states if their connectivity is suitably chosen to represent relevant higher order correlations, even if the model remains low order.
\item The number of higher-order terms appearing on the right of~(\ref{eq:momentClosure}) grows with the order on the left. This problem is compounded if species labels are included. Hidden layers and a restriction on the number of species $M^\prime$ allowed to occupy hidden units may be used to approximate such higher order interactions with fewer parameters.
\end{enumerate}

It is generally difficult to choose the optimal close approximation, i.e. to know which moments are relevant to the time-evolution of a given dataset.
A key advantage of the present approach is that the connectivity of the hidden layers may be chosen based on the differential equations derived from the chemical master equation.
For example, consider to the bimolecular annihilation system~(\ref{eq:momentClosure}): if the goal is to accurately model the mean number of particles, then the right hand side of~(\ref{eq:momentClosure}) shows that the nearest-neighbor moment is relevant to the time evolution. The graphical model of the reduced system could therefore introduce a hidden unit for every pair of neighboring lattice sites, with corresponding energy function:
\begin{equation}
\begin{split}
& E ({\boldsymbol v}, {\boldsymbol h} , b(t), \weightVH(t), \biasH(t) ) = - b(t) \sum_{i=1}^N v_i \\
& - \biasH(t) \sum_{j=1}^{N-1} h_j - \weightVH(t) \sum_{i=1}^{N} \sum_{j=i-1,i} v_i h_{j} ,
\end{split}
\label{eq:model21}
\end{equation}
and differential equation model:
\begin{equation}
\begin{split}
\dot{b} =& F_{b}(b,b^\prime,W; \ubf_{b}) , \\
\dot{b}^\prime =& F_{b^\prime}(b,b^\prime,W; \ubf_{b^\prime}) , \\
\dot{W} =& F_{W}(b,b^\prime,W; \ubf_{W}) .
\end{split}
\label{eq:model22}
\end{equation}
The corresponding graphical model is shown in Figure~\ref{fig:bimolModels}(a,c). 

\begin{figure*}[!t]
	\centering
	\includegraphics[width=0.75\textwidth]{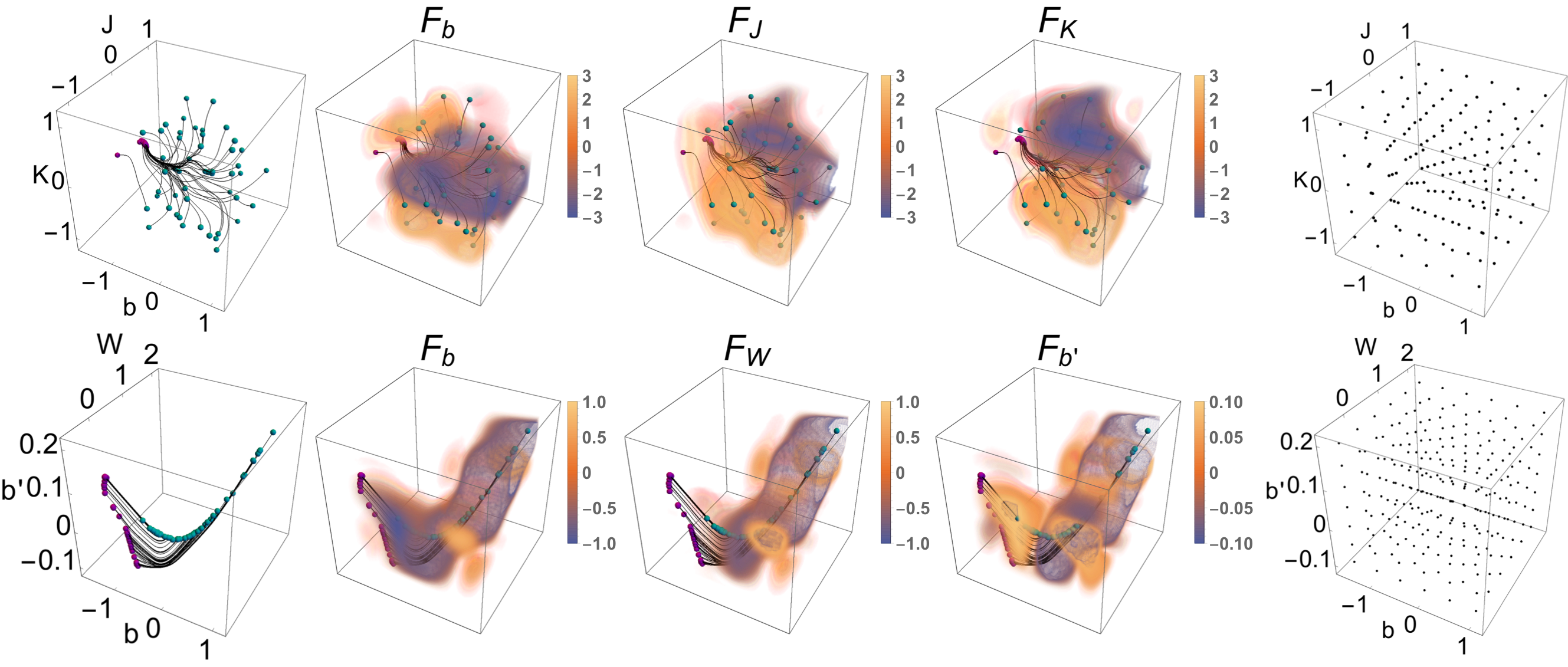}
	\caption{
    \textit{Top row:} Learned time-evolution functions for the fully visible model~(\ref{eq:model12}), using the $Q_3$, $C_1$ finite element parameterization~(\ref{eq:q3}) with cells of size $0.5\times 0.5\times 0.5$ in $(b,J,K)$. 
    Left panel: Training set of initial points $(b,J,K)$ (cyan) sampled evenly in $[-1,1]$. Stochastic simulations for each initial point are used as training data (learned trajectories shown in black, endpoints in magenta). Middle three panels: the time evolution functions learned, where the heat map indicates the value of $F_\gamma$ in~(\ref{eq:model12}). Right panel: vertices of the finite element cells used.
        \textit{Bottom row:} Hidden layer model~(\ref{eq:model22}) and parameterization~(\ref{eq:q3}) with cells of size $0.5\times0.5\times0.05$ in $(b,W,b^\prime)$. Initial points are generated by BM learning applied to the points of the visible model.
        Note that the coefficients corresponding to the other seven degrees of freedom at each vertex are also learned (not shown), i.e. the first derivatives in each parameter.
    }
    \label{fig:bimolBFs}
\end{figure*}

\begin{figure*}[!t]
	\centering
	\includegraphics[width=0.75\textwidth]{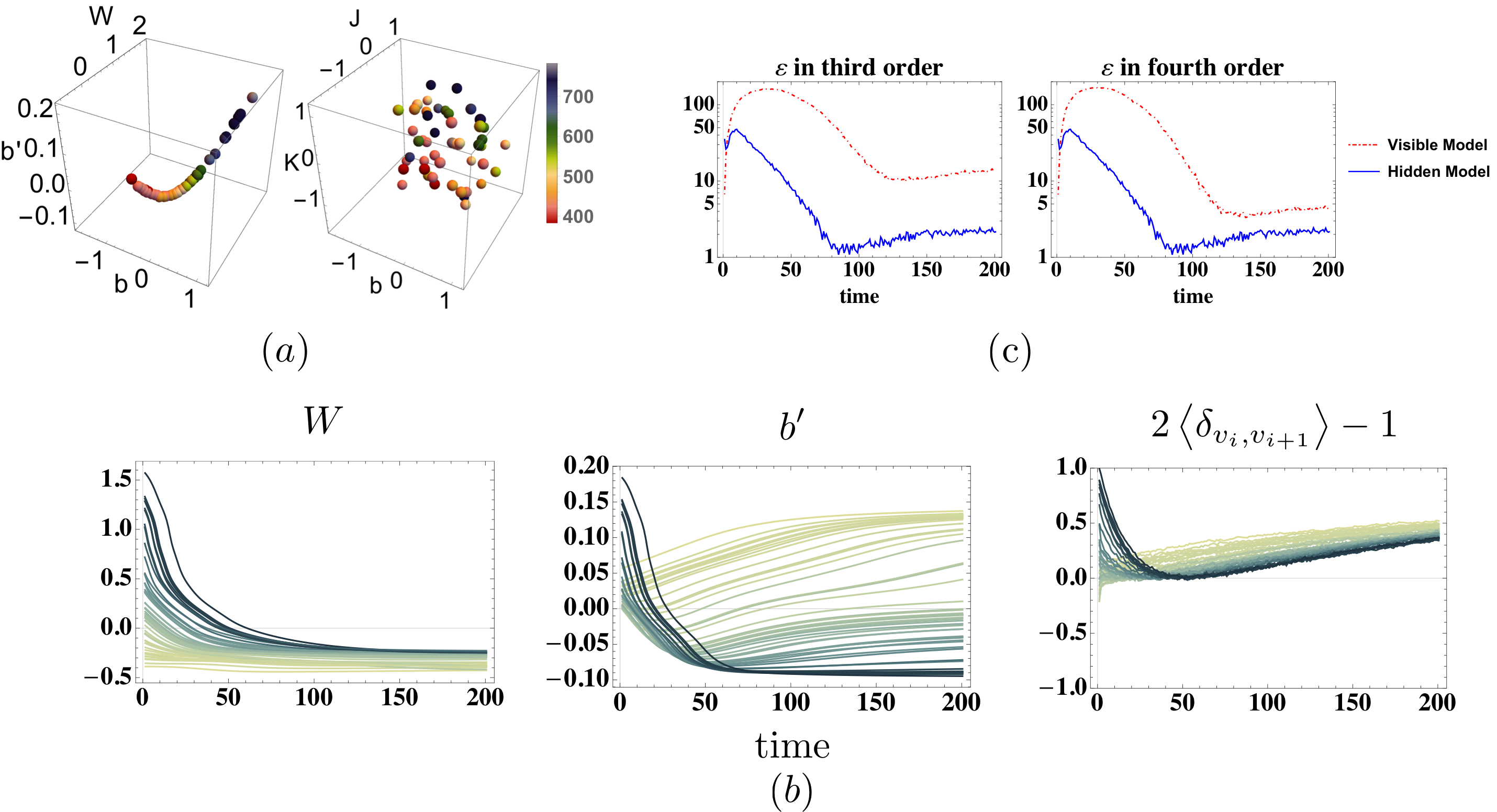}
	\caption{
	(a) Nearest neighbor moment $\left \langle \sum_i v_i v_{i+1} \right \rangle$ of the two models. The more compact representation learned by the hidden layer model (left) captures low range spatial correlations, while the fully visible model (right) shows no apparent organization.
(b) The parameters $W$ and $b^\prime$ for the hidden layer model for the 50 initial conditions ($b$ is monotonically decreasing for all trajectories). The learned parameters encode the spatial correlation $2 \langle v_i v_{i+1} \rangle$ shown on the right. This shows the moment closure approximation learned by the reduced model (see text).
(c) RMSE in the third order moment $\langle \sum_i v_i v_{i+1} v_{i+2} \rangle$ and fourth order moment $\langle \sum_i v_i v_{i+1} v_{i+2} v_{i+3} \rangle$, calculated from a set of test trajectories (not shown). Both models reproduce the observables with reasonable accuracy, however, the error in the hidden layer model is lower due to the more compact representation learned.
    }
    \label{fig:bimolErrors}
\end{figure*}

The time-evolution functions for~(\ref{eq:model12}) and~(\ref{eq:model22}) are learned using Algorithm~1 and compared in Figure~\ref{fig:bimolBFs}. For the visible model, cells of size $0.5\times0.5\times0.5$ in $(b,J,K)$ are used, and for the hidden layer model cells of size $0.5\times 0.5 \times 0.05$ in $(b,W,b^\prime)$, as shown in Figure~\ref{fig:bimolBFs}.

As training data, 50 points $(b,J,K)$ are sampled evenly over $(b,J,K) \in [-1,1]^3$. Each point corresponds to an initial distribution~(\ref{eq:model11}), from each of which 50 lattices of length $N=1000$ are sampled (top left panel of Fig~\ref{fig:bimolBFs}). The corresponding initial conditions in $(b,W,b^\prime)$ space are learned separately using the BM learning algorithm (bottom left panel of Fig~\ref{fig:bimolBFs}). Each lattice is simulated for $200$ timesteps of size $\Delta t = 0.01$ with $p_r = 0.01$, as shown in Figure~\ref{fig:bimolModels}(e). These trajectories are pooled for Algorithm~1. Note that a single set of parameter vectors $\{ \ubf \}$ in (\ref{eq:model12},\ref{eq:model22}) is learned, i.e. the parameter vectors are shared among trajectories from all initial conditions.

For the fully visible model, asleep phase moments are estimated by running a Gibbs sampler for a single step. Similarly, for the hidden model, awake and asleep phase moments are estimated by a single step of contrastive divergence, i.e. CD-1.
The parameters to Algorithm~1 are learning rate $\lambda=1$ for $200$ optimization steps for both models. 

The time integral in the action~(\ref{eq:actionDiscrete}) can lead to undesired extrema, e.g. for periodic trajectories. We use an on-line algorithmic solution is to shift the limits of integration in~(\ref{eq:sensitivity}) as new data is available:
\begin{equation}
\frac{d S}{d u_{k,i}} = \int_{\tau}^{\tau + \Delta \tau} dt \; \frac{\partial F_k (\thetabf(t) ; \ubf_k)}{\partial u_{k,i}} \phi_k(t)
\label{eq:sensitivityTau}
\end{equation}
where $\Delta \tau$ is fixed, and $\tau$ is gradually incremented $t_0 \leq \tau \leq t_f - \Delta \tau$. In this case, the PDE constraint~(\ref{eq:pdeParameterized}) is solved from $t_0$ to $\tau$, decreasing the size of the trajectories early in the training. Further, the adjoint system~(\ref{eq:adjoint}) only has to be solved backwards from $\phi(\tau + \Delta \tau) = 0$ to $\phi(\tau)$, which also controls the magnitude of the update steps as the length of the trajectory grows, allowing a constant learning rate to be used. For the annihilation system, we found that fixing $\Delta \tau = 5$ timesteps and shifting $\tau \rightarrow \tau+1$ every $2$ optimization steps gave fast convergence.

Figure~\ref{fig:bimolBFs} shows the learned time-evolution functions and trajectories of the training data. For the visible model, these show an expected symmetric structure. As particles diffuse and nearest-neighbor (NN) and next-nearest-neighbor (NNN) moments decay, $F_J$ and $F_K$ force $J,K \rightarrow 0$ everywhere, while the bias term tends to negative infinity. The representation learned by the hidden layer model is more compact. Figure~\ref{fig:bimolErrors}(a) shows the nearest neighbor moment $\left \langle \sum_i v_i v_{i+1} \right \rangle$ overlaid onto the initial conditions, showing an almost monotonic organization from low to high values by which the model can distinguish these states (no organization is apparent in the visible model). Figure~\ref{fig:bimolErrors}(b) shows the learned parameter trajectories: $b$ monotonically decreases (not shown), $W$ asymptotically approaches a negative value, and $b^\prime$ either increases monotonically or initially decreases before increasing again. This division corresponds to the decay of spatial correlations $2 \langle v_i v_{i+1} \rangle -1$ (such that $1$ corresponds to a fully correlated lattice, and $-1$ to a fully anti-correlated lattice), also shown in Figure~\ref{fig:bimolErrors}(b). The two types of trajectories of $b^\prime$ have a clear correspondence to two types of trajectories in the correlation function, and the separation is visible in $F_{b^\prime}$ in the negative and positive regimes. We conclude that the moment closure approximation learned by the model therefore captures relevant low range spatial correlations to approximate the right hand sides of the moment equations~(\ref{eq:momentClosure}) identified from the CME.

To assess the accuracy of the reduced models, we generate a test set of points $(b,J,K)$ and the learn the corresponding points $(b,W,b^\prime)$ as before. These are evolved in time using the learned DE systems~(\ref{eq:model12},\ref{eq:model22}).
Define $\varepsilon(t) = \sqrt{ \langle ( \mu(t) - \tilde{\mu}(t) )^2 \rangle }$ as the root mean square error (RMSE) between some moments of the reduced model $\tilde{\mu}$ and the stochastic simulations $\mu$, where the moments are approximated by averaging over $50$ samples. Figure~\ref{fig:bimolErrors}(c) shows the RMSE for the third order moment $\langle \sum_i v_i v_{i+1} v_{i+2} \rangle$ and fourth order moment $\langle \sum_i v_i v_{i+1} v_{i+2} v_{i+3} \rangle$. 
Both models have relatively low error in reproducing the observables, however, the error in the hidden layer model is lower than in the visible model. This is because the representation learned by the hidden layer model is more compact, in that states initially distributed uniformly in $(b,J,K)$ space are mapped to an approximately 1D curve in $(b,W,b^\prime)$ space. Yet higher accuracies may be possible by further tailoring that parameterizations of the differential equations from the cubic finite elements used here.


\subsection{Learning the R{\"o}ssler oscillator} \label{sec:4.2}


\begin{figure}[t]
	\centering
	\includegraphics[width=0.5\textwidth]{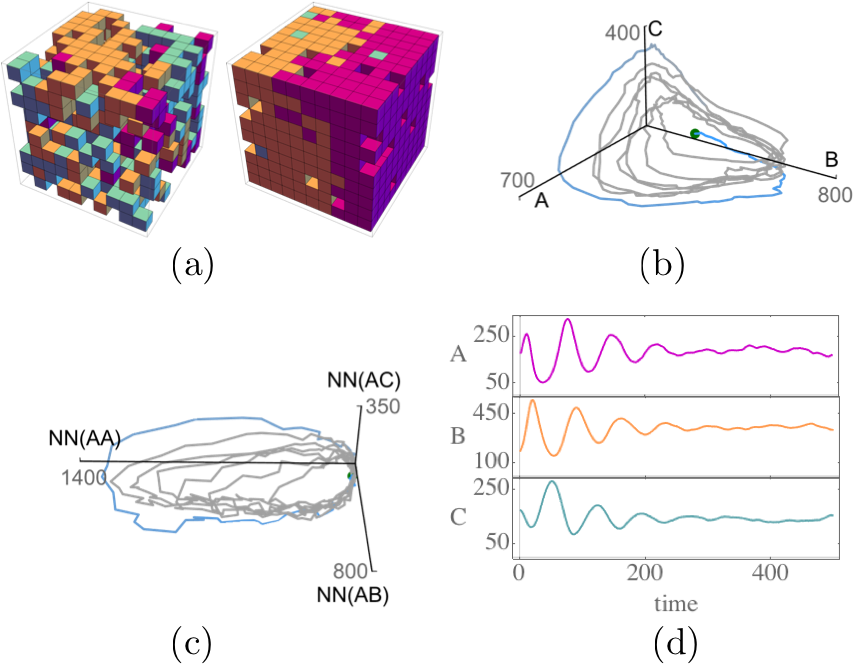}
	\caption{
    R{\"o}ssler oscillator on a 3D lattice. 
    (a) Snapshots of a stochastic simulation on a $10 \times 10 \times 10$ lattice ($A,B,C$ in pink, orange, cyan). 
    (b) Moments from a single simulation over $500$ timesteps, producing a stochastic version of the characteristic attractor of the well-known deterministic model. 
    (c) Nearest neighbor moments in the simulation of~(b) show similar structure.
    (d) Relaxation to a stationary distribution, indicated by the convergence of the means from averaging over $300$ stochastic simulations.
    }
	\label{fig:rossler}
\end{figure}

\begin{figure}[t]
	\centering
	\includegraphics[width=0.4\textwidth]{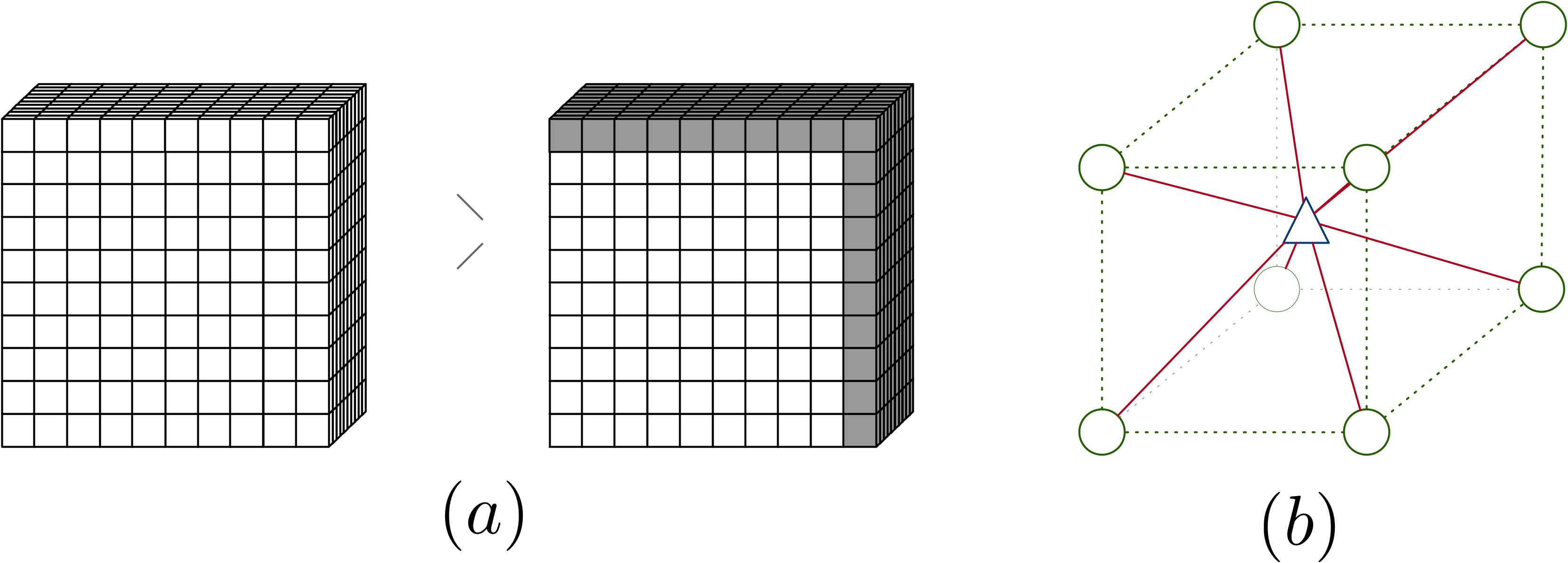}
	\caption{(a) Graph to learn for the R{\"o}ssler oscillator. The lattice on the left corresponds to the visible layer, equivalent to the $10\times10\times10$ cube in Figure~\ref{fig:rossler}; the right corresponds to the hidden layer. 
	Gray units in the hidden layer denote those units which implement periodic boundary conditions to the visible layer. (b) Connectivity of hidden layer. Each cube of 8 neighboring units in the visible layer (green circles) is connected to a single unit (blue triangle) in the hidden layer (connections shown in red), resembling a body-centred cubic structure. Biases for the units are not shown.
	}
	\label{fig:rosslerGraph}
\end{figure}



\begin{figure}[t]
	\centering
	\includegraphics[width=0.4\textwidth]{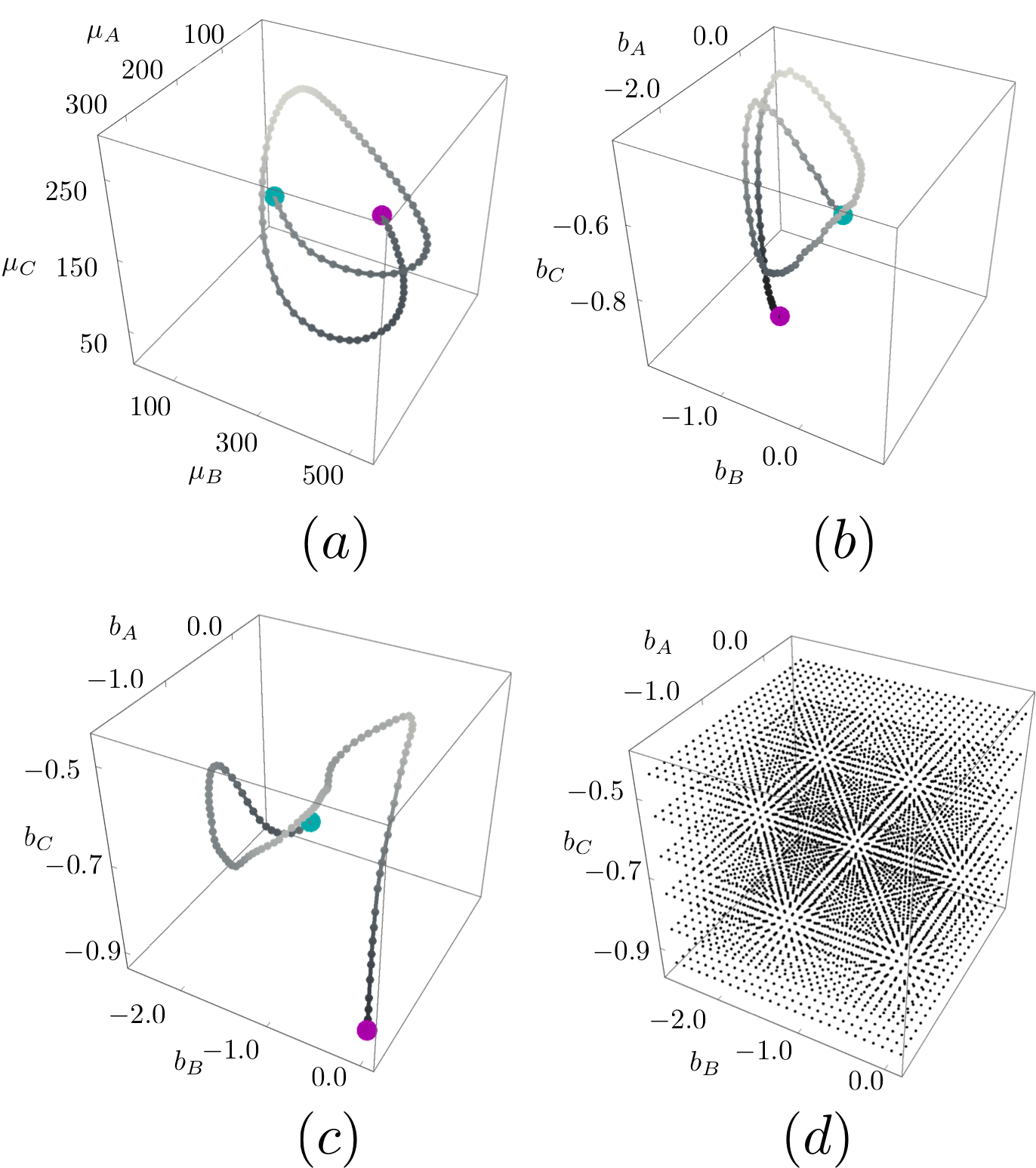}
	\caption{(a) The first $100$ timesteps of the mean number of $A,B,C$ in the R{\"o}ssler oscillator system. (b) Interaction parameters for a MaxEnt model constrained on the moments in (a) given by equation~(\ref{eq:simpleMaxEnt}). (c) The learned trajectory of~(\ref{eq:rosslerModel1}) in $(b_A,b_B,b_C)$-space, with initial condition $(-\ln(2),-\ln(2),-\ln(2))$. The bias parameters have been tuned to control both the means and spatial correlations, together with the weights (not shown). Gray scale value indicates $b_C$ component for clarity, scaled from dark ($\text{min}(b_C)$) to light ($\text{max}(b_C)$). Initial point is shown in cyan, and endpoint in magenta. (d) Vertices of the finite element cells of side length $0.1$ used to parameterize the differential equations~(\ref{eq:rosslerModel2}).
	}
	\label{fig:rosslerTrajCompare}
\end{figure}


\begin{figure}[t]
	\centering
	\includegraphics[width=0.35\textwidth]{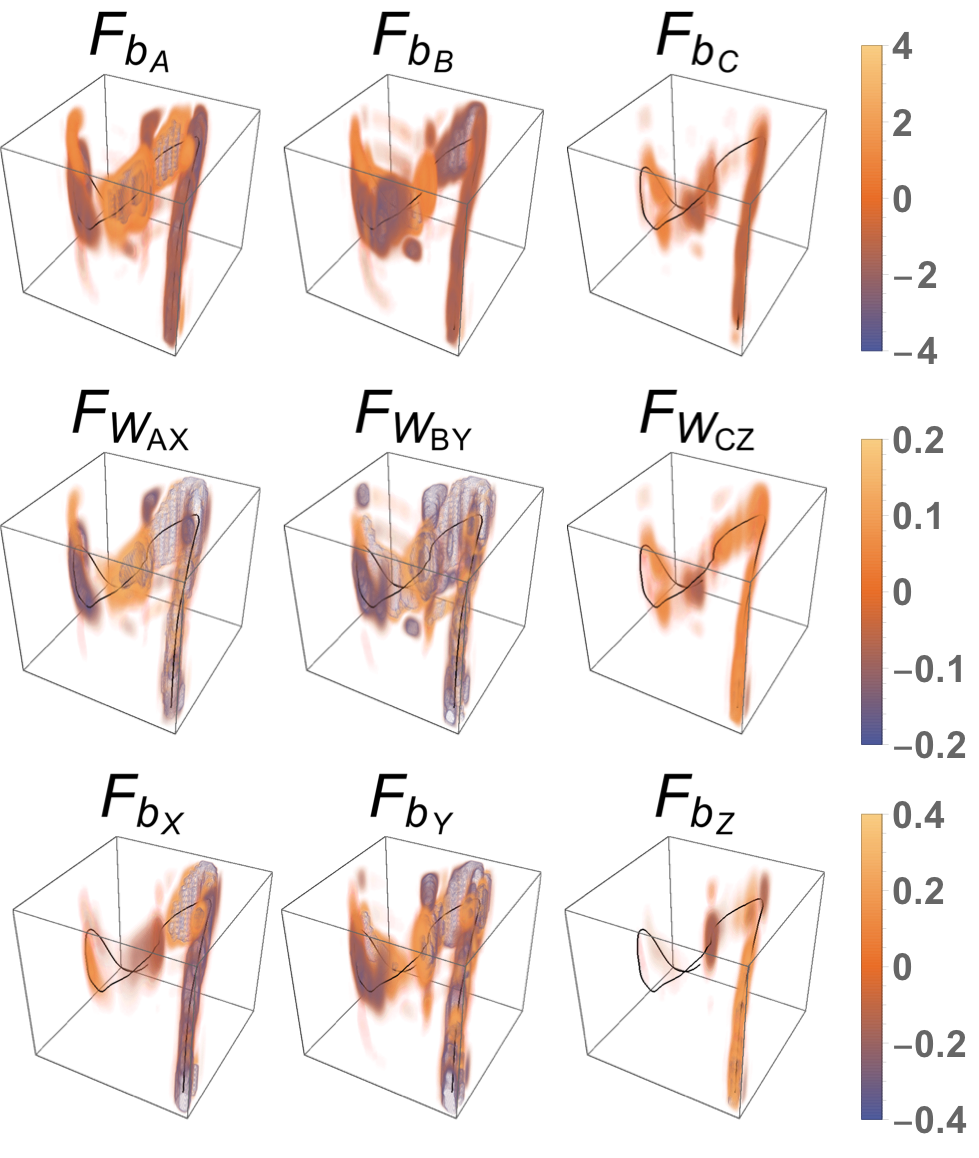}
	\caption{Learned time-evolution functions~(\ref{eq:rosslerModel2}) in $(b_A,b_B,b_C)$-space (see Figure~\ref{fig:rosslerTrajCompare}(d) for the vertices used), and the resulting trajectory in black (see Figure~\ref{fig:rosslerTrajCompare}(c)).
    }
	\label{fig:rosslerBFs}
\end{figure}


\begin{figure*}[t]
	\centering
	\includegraphics[width=0.8\textwidth]{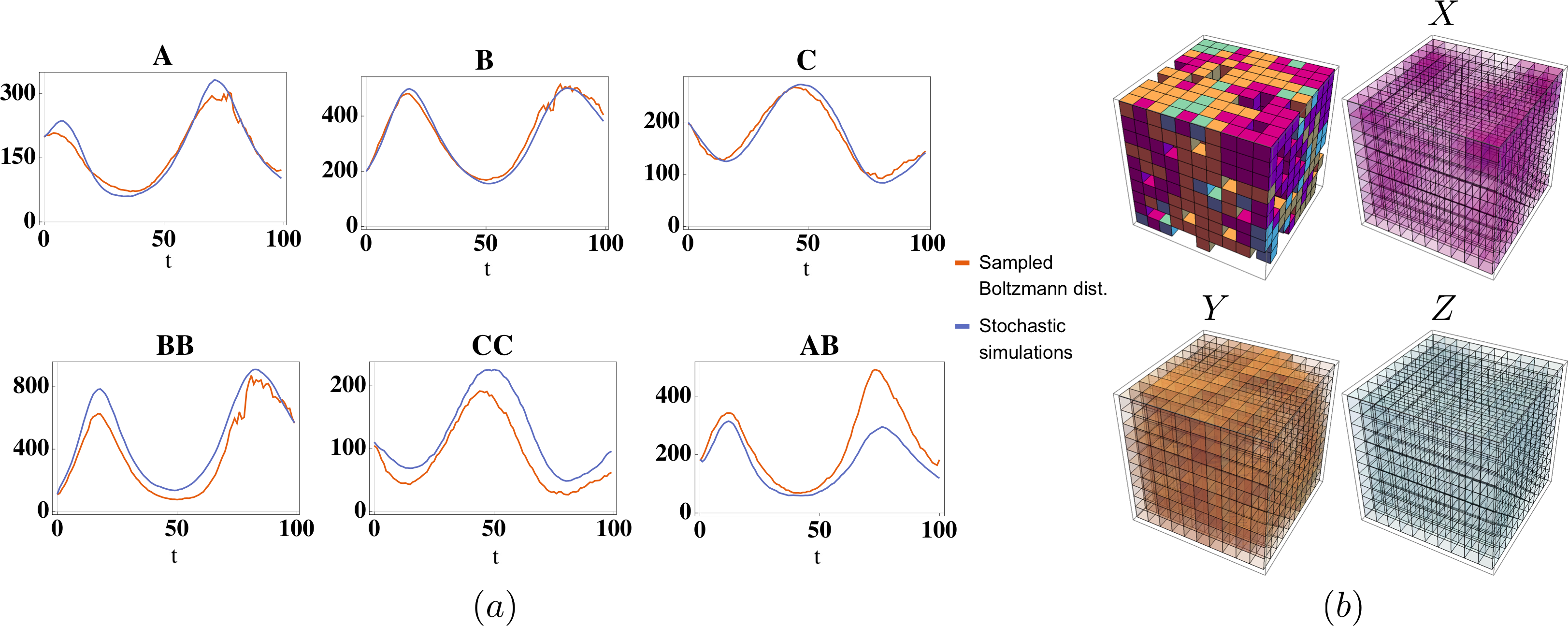}
	\caption{
		(a) Example of correlations learned by the reduced model compared to stochastic simulations, obtained by sampling over $100$ samples. Top row: mean number of $A,B,C$ particles. Bottom: neighboring pairs of $(B,B),(C,C),$ and $(A,B)$. Short range spatial correlations relevant to the moment equations~(\ref{eq:deModel}) are reasonably approximated due to the chosen connectivity.
		(a) Sampled state $\boldsymbol{V}$ from the learned model (top left), and the activated hidden layer probabilities $\pt(\boldsymbol{H} | \boldsymbol{V})$ at timepoint $20$. After training, the hidden layers coarse grain nearest neighbors in the visible layer.
    }
	\label{fig:rosslerResult}
\end{figure*}


The Williamowski-R{\"o}ssler oscillator system~\cite{
williamowski_1980} is a chemical version of a spiral oscillator in three species. The 
original 
formulation requires additional species that are 
fixed 
at constant concentration. We follow recent work~\cite{giovanni_2016} on a volume-excluding version where these constraints are incorporated into pseudo-first order reaction rates. The oscillator for the species $A,B,C$ is dictated by the reaction system:
\begin{equation}
\begin{split}
& A \xrightleftharpoons[p_1]{k_1} 2 A 
\qquad A + B \xrightarrow{p_2} 2 B
\qquad A + C \xrightarrow{p_3} \oset \\ 
& \hspace{12mm} \qquad B \xrightarrow{k_2} \oset
\qquad C \xrightleftharpoons[p_4]{k_3} 2 C
\end{split}
\end{equation}
where the unimolecular reaction rates used are $k_1 = 30, k_2 = 10, k_3 = 16.5$ (arbitrary units), and the probabilities for bimolecular reactions are $p_1 = 0.1, p_2 = 0.4, p_3 = 0.24, p_4 = 0.36$. We simulate this system on a 3D lattice of size $10\times10\times10$ sites in the single occupancy limit as before. Figure~\ref{fig:rossler} shows snapshots of such a stochastic simulation. Panel~(b) in particular shows the characteristic shape of the R{\"o}ssler oscillator, with further structures evident in higher order moments shown in (c). A snapshot of the spatial waves that occur during transitions between $A,B$ and $C$-dominated regimes is shown in panel~(a). 

The time evolution of the mean number of particles in $A,B,C$, denoted by $\mu_\alpha$, is related to the number of nearest neighbors, denoted by $\Delta_{\alpha \beta}$, as follows:
\begin{equation}
\begin{split}
\frac{d}{dt} \mu_A &= k_1 \mu_A - \kappa_1 \Delta_{AA} - \kappa_2 \Delta_{AB} - \kappa_3 \Delta_{AC} , \\
\frac{d}{dt} \mu_B &= \kappa_2 \Delta_{AB} - k_2 \mu_B , \\
\frac{d}{dt} \mu_C &= - \kappa_3 \Delta_{AC} + k_3 \mu_C - \kappa_4 \Delta_{CC} ,
\end{split}
\label{eq:deModel}
\end{equation}
where $\kappa_1,\kappa_2,\kappa_3,\kappa_4$ are the reaction rates for the bimolecular reactions specified by probabilities $p_1,p_2,p_3,p_4$ above. As previously, this system is not closed, such that two close initial states in Figure~\ref{fig:rossler}(b) will diverge over their long term time-evolution. The challenge for the latent variables in the reduced differential equation model is to incorporate relevant higher order correlations to separate states which are close in their lower order moments. 

As in Section~\ref{sec:4.1}, let the visible part of the graph be the lattice of Figure~\ref{fig:rossler}(a). For the hidden layer, we choose a connectivity that coarse-grains the visible lattice by one unit in each spatial dimension
as shown in Figure~\ref{fig:rosslerGraph}. Note that the hidden layer is also of size $10\times10\times10$ units that implement periodic boundary conditions. The visible layer of the graph is multinomial in one of $\{A,B,C,\oset\}$, and similarly the hidden layer in $\{ X, Y, Z, \oset \}$.
The corresponding energy model is:
\begin{equation}
\begin{split}
& E({\boldsymbol V}, {\boldsymbol H}, {\boldsymbol \theta}(t) ) = \\
& - \sum_{i} \, \sum_{\alpha\in\{A,B,C\}} b_\alpha v_{i,\alpha}
- \sum_{j} \, \sum_{\alpha \in \{X,Y,Z \}} b_{\alpha} h_{j,\alpha} \\
& - \sum_{\{ i,j \}} \, \Big ( W_{AX} v_{i,A} h_{j,X} + W_{BY} v_{i,B} h_{j,Y}  + W_{CZ} v_{i,C} h_{j,Z} \Big )
,
\end{split}
\label{eq:rosslerModel1}
\end{equation}
where ${\boldsymbol H}$ refers to the hidden layer, and the sum over $\{i,j\}$ implements the connectivity shown in Figure~\ref{fig:rosslerGraph}, and
\begin{equation}
\dot{\gamma} = F_\gamma (b_A, b_B, b_C; \ubf_\gamma ) 
\label{eq:rosslerModel2}
\end{equation}
for $\gamma \in \{ b_A, b_B, b_C, W_{AX}, W_{BY}, W_{CZ}, b_X, b_Y, b_Z \}$.
The right hand side of the differential equation is parameterized~(\ref{eq:q3}) by cubic $C_1$ finite elements as before. To reduce the complexity of the model, we have purposefully omitted interactions $W_{AY},W_{AZ},W_{BX},W_{BZ},W_{CX},W_{CY}$. With this choice, the latent species $X$ coarse grains the visible species $A$, and similarly for $Y,B$ and $C,Z$. Note that all differential equation models share the same domain in $(b_A, b_B, b_C)$ space. Note that while the biases $h_A,h_B,h_C$ are the Lagrange multipliers corresponding to the constraints for the number of particles of each species, through the energy function~(\ref{eq:rosslerModel1}) both the biases and weights together also control all spatial correlations of the model.

Stochastic simulations are generated from an initial state with $b_A = b_B = b_C = - \ln(2)$, $W_{AX} = W_{BY} = W_{CZ} = W_{XY} = W_{YZ} = 0$, and $b_X = b_Y = b_Z = - \ln(1/7)$. By setting the initial weights to zero, this is the MaxEnt state given that the number of particles is $\mu_A = \mu_B = \mu_C = 200$, since with zero weight:
\begin{equation}
\mu_\alpha = 1000 \times \frac{
e^{b_\alpha} 
}{
1 + \sum_{\beta=A,B,C} e^{b_{\beta}}
}
\quad
\text{for } \alpha \in \{ A,B,C \}
\label{eq:simpleMaxEnt}
\end{equation}
where the factor $1000$ results from summing over all visible sites. With zero weight, the choice for the initial hidden layer bias is free - by choosing to set it to 
$-\ln(1/7)$, we are setting the target sparsity to approximately half of that of the visible layer with approximately $100$ particles of each species as given by~(\ref{eq:simpleMaxEnt}).
Simulations are run for 500 timesteps of size $\Delta t = 0.01$. Figure~\ref{fig:rossler}(d) shows the relaxation of the distribution to equilibrium~\cite{anishchenko_2004}.

For training, we use Algorithm~1 with learning rate $\lambda=0.05$ for the weights and $\lambda = 0.8$ for the biases for $10000$ optimization steps. To estimate the awake phase moments, we sample $\pt(\boldsymbol{H}=1 | \boldsymbol{V})$ for each sample in a batch size of $\eta=5$, where $\boldsymbol{V}$ is a data vector. To estimate the asleep phase moments, we alternate between sampling $\pt(\boldsymbol{H}^{(r)}=1 | \boldsymbol{V}^{(r)})$ and $\pt(\boldsymbol{V}^{(r)}=1 | \boldsymbol{H}^{(r-1)})$ for $r=1,\dots,10$ steps, starting from a random configuration $\boldsymbol{V}^{(0)}$. Alternatively, we also found fast convergence using $k=10$ steps of contrastive divergence (CD), as well as using persistent CD. To reduce the noise in the estimates, we use as is common raw probabilities instead of multinomial states for the hidden units when estimating both the awake and asleep phase moments.

As before, we use the online variant~(\ref{eq:sensitivityTau}) of Algorithm~1 where the limits of integration are shifted during training, with window size $\Delta \tau = 10$, and $\tau$ is gradually incremented $\tau \rightarrow \tau +1$ every 100 optimization steps.
To learn smooth trajectories and avoid jumps in the learned differential equation model, each timestep is divided into $10$ substeps when solving the differential equations~(\ref{eq:rosslerModel1},\ref{eq:rosslerModel2}). 

Figure~\ref{fig:rosslerBFs} shows the learned time evolution functions for the R{\"o}ssler oscillator over the first $100$ timesteps. The side length of the cubic finite elements used was $0.1$ on all sides, centered at the initial condition, as shown in Figure~\ref{fig:rosslerTrajCompare}(d). 
We compare the learned trajectories to a simplified MaxEnt model in Figure~\ref{fig:rosslerTrajCompare}(a)-(c). Panel~(a) shows the mean number of particles over the first $100$ timesteps, as in Figure~\ref{fig:rossler}(d). Panel~(b) transforms these points to the parameters~$(b_A,b_B,b_C)$ of a simple MaxEnt model constrained on these lowest order moments as given by~(\ref{eq:simpleMaxEnt}). Panel~(c) shows the learned model~(\ref{eq:rosslerModel2}), where the biases now control both the means and spatial correlations together with the weights. The trajectory no longer resembles a periodic trajectory, having learned to separate close states in panel~(b).

The agreement between the stochastic simulations and reconstructed observables is shown in Figure~\ref{fig:rosslerResult}(a). At each timepoint, $100$ samples are drawn from the reduced model by running $25$ steps of CD sampling, starting from a random configuration. Nearest neighbors, which determine the time evolution of the means in~(\ref{eq:deModel}) are reasonably approximated, primarily due to the connectivity chosen in Figure~\ref{fig:rosslerGraph}.

Figure~\ref{fig:rosslerResult}(b) shows a sampled state $\boldsymbol{V}$ from the learned model, and the activated hidden layer probabilities $\pt(\boldsymbol{H} | \boldsymbol{V})$ at timepoint $20$. With the learned parameters, the hidden units coarse grain nearest neighbors in the lattice, as needed to approximate the right hand side of~(\ref{eq:deModel}). A deeper network such as a deep Boltzmann machine (DBM) may approximate yet higher spatial correlations, and can therefore be used to close differential equation systems depending on higher order moments.


\section{Discussion} \label{sec:5}


We have presented a learning problem for spatiotemporal distributions that estimates differential equation systems controlling a time-varying Boltzmann distribution. 
The ability to enforce a reduced physical model to be estimated makes the method interesting for many modeling applications, including chemical kinetics as presented here.
Mapping to a differential equation model can be likewise be useful for engineering applications, allowing constraints to be efficiently introduced into BM learning as discussed in Section~\ref{sec:3.1}.

The moment closure approximation presented in Section~\ref{sec:2} is broadly applicable due to the use latent variables that can be trained to capture relevant higher order correlations, rather than deciding a priori what correlations to include as in typical closure schemes. Minimizing the KL divergence between the reduced and true models at all times is closely related to entropic matching, but differs by the introduction of a differential equation system. We also make the connection to spatially continuous reaction systems explicit.

The finite element parameterization is similar to the unsupervised learning setting of RBMs in the sense that it is independent of the system under consideration. For deeper architectures such as DBMs as discussed in Section~\ref{sec:4.2}, recycling the same time-evolution functions across multiple layers may be effective, similar to convolution layers in convolutional neural networks. Factoring weights has also been used effectively in deep learning~\cite{ranzato_2010}, and may similarly reduce the computational burden here. The main advantage of the current DE formalism, however, is to use a parameterization~(\ref{eq:pde}) that enforces a physically relevant model. 

A popular alternative class of generative models to RBMs are variational autoencoders (VAEs). An adaptation of the proposed method may be possible for these models - however, the main advantages of the current RBM framework is that the form of the energy function can be used interpret the reduced model~\cite{ernst_2018}, and that the distribution over the latent variables is not chosen as in VAEs (typically a standard normal distribution), but rather learned from data.

A closely related problem to model reduction is the problem of data assimilation, where noisy measurements and an incomplete model for the dynamics are combined to estimate the true state of the system and unknown parameters in the model~\cite{abarbanel_book}. Model reduction methods complement the data assimilation problem by replacing the physical model with a reduced one which can increase the efficiency of data assimilation methods. 

We view the present work as progress toward linking models across scales in biology~\cite{mjolsness_2018}. Reaction-diffusion systems illustrate many of the common problems in this field. While much machinery (CME or field-theoretic methods) exists to formulate problems for observables, their solution is non-trivial in most applications. Even without analytic challenges such as moment closure, the numerical solution of PDE systems is difficult for systems with high spatial organization, or where interactions with other scales (e.g. molecular dynamics) or physics (e.g. electrodiffusion) become relevant. Learning reduced models in the form of spatial dynamic Boltzmann distributions may abstract much of these non-trivial interactions.


\begin{acknowledgments}

This work was supported by NIH grants R01HD073179 and USAF/DARPA FA8750-14-C-0011 (E.M.) and NIH P41-GM103712 and AFOSR MURI FA9550-18-1-0051 (O.K.E., T.B., T.S.).

\end{acknowledgments}


\appendix

\section{Formal solution for the adjoint system} \label{app:1}


The connection between~(\ref{eq:varProblemGeneral}) and~(\ref{eq:varProblemAutonomous}) can be made more explicitly. A differential equation system for the perturbations $\delta \nu_k(\alb_\anglink,\xb_\anglink,t)$ in~(\ref{eq:varProblemGeneral}) can be derived by linearizing the differential equation around a particular solution~\cite{ernst_2018,gamkrelidze_book}. For the autonomous system~(\ref{eq:pdeAutonomous}), this leads to the linear ODE system:
\begin{equation}
\frac{d}{dt} \delta {\boldsymbol \nu}(\alb,\xb,t)
=
\delta {\boldsymbol F} (\alb,\xb,t)
+
G(\alb,\xb,t)
\delta {\boldsymbol \nu}(\alb,\xb,t)
\label{eq:linearVar}
\end{equation}
with some given initial condition $\delta {\boldsymbol \nu}(\alb,\xb,t_0) = \delta {\boldsymbol \eta}(\alb,\xb)$. Here we have used the vector notation introduced in Section~\ref{sec:2.2}.

Let the homogenous part of this system:
\begin{equation}
\frac{d}{dt} \delta {\boldsymbol \nu}(\alb,\xb,t)
=
G(\alb,\xb,t)
\delta {\boldsymbol \nu}(\alb,\xb,t)
\end{equation}
have solution given by the non-singular fundamental matrix $A(\alb,\xb,t)$. Then~(\ref{eq:linearVar}) has as formal solution
\begin{equation}
\begin{split}
\delta {\boldsymbol \nu} (\alb,\xb,t) 
= & A(\alb,\xb,t) \Bigg (
\delta {\boldsymbol \eta}(\alb,\xb) \\
& +
\int_{t_0}^{t} d\tp \; A^{-1}(\alb,\xb,\tp) \delta {\boldsymbol F} (\alb,\xb,\tp)
\Bigg )
\end{split}
\end{equation}
which substituted into~(\ref{eq:varProblemGeneral}) gives:
\begin{widetext}
\begin{equation}
\delta S = \int_{t_0}^{t_f} dt \; \sum_{n=0}^\infty \sum_{\alb} \int d\xb \;  
\Delta {\boldsymbol \mu}^\intercal (\alb,\xb,t) 
A(\alb,\xb,t) 
\left (
\delta {\boldsymbol \eta}(\alb,\xb) +
\int_{t_0}^{t} d\tp \; A^{-1}(\alb,\xb,\tp) \delta {\boldsymbol F} (\alb,\xb,\tp)
\right ) 
= 0
\label{eq:varProblemSub}
\end{equation}
\end{widetext}
where $\Delta {\boldsymbol \mu}^\intercal (t)$ is the vector with components~(\ref{eq:momentDiff}). Applying integration by parts on the term in parentheses to move the integral over time gives
\begin{widetext}
\begin{equation}
\begin{split}
&
\left ( 
\delta {\boldsymbol \eta} (\alb,\xb) +
\int_{t_0}^{t_f} d\tp \; A^{-1}(\alb,\xb,\tp) \delta {\boldsymbol F} (\alb,\xb,\tp)
\right )
\left (
\int_{t_0}^t d\tp \; \Delta {\boldsymbol \mu}^\intercal (\alb,\xb,\tp) A(\alb,\xb,\tp)
\right )
\Bigg |_{t=t_0}^{t_f} \\
&-
\int_{t_0}^{t_f} dt \; \int_{t_0}^t d\tp \; \Delta {\boldsymbol \mu}^\intercal (\alb,\xb,\tp) A(\alb,\xb,\tp) A^{-1}(\alb,\xb,t) \delta {\boldsymbol F} (\alb,\xb,t)
\end{split}
\label{eq:bdry}
\end{equation}
\end{widetext}
where the adjoint functions ${\boldsymbol \zeta}(t)$ can be identified as:
\begin{equation}
{\boldsymbol \zeta}^\intercal(\alb,\xb,t) = \int_{t_0}^t d\tp \; \Delta {\boldsymbol \mu}^\intercal (\alb,\xb,\tp) A(\alb,\xb,\tp) A^{-1}(\alb,\xb,t)
\end{equation}
By choosing the adjoint functions to satisfy the boundary condition ${\boldsymbol \zeta}(\alb,\xb,t_f) = 0$, the boundary term in~(\ref{eq:bdry}) vanishes and we obtain the previous result~(\ref{eq:varProblem}).



\bibliographystyle{apsrev4-1}
\bibliography{bibliography}

\end{document}